\documentclass[showpacs,twocolumn,aps,prb]{revtex4-1}
\usepackage{graphicx}
\usepackage{dcolumn}
\usepackage{bm}
\usepackage{amsmath}
\usepackage{amssymb}
\usepackage[usenames,dvipsnames]{xcolor}
\usepackage{color}

\begin{document}
\title{Unidirectional Seebeck effect in magnetic topological insulators}

\author{Xiao-Qin Yu$^{1}$}
\email{yuxiaoqin@hnu.edu.cn}

\author{Zhen-Gang Zhu$^{2,3,4}$}
\email{zgzhu@ucas.ac.cn}

\author{Gang Su$^{3,4,5}$}
\email{gsu@ucas.ac.cn}

\affiliation{
$^{1}$ School of Physics and Electronics, Hunan University, Changsha 410082, China.\\
$^{2}$ School of Electronic, Electrical and Communication Engineering, University of Chinese Academy of
Sciences, Beijing 100049, China. \\
$^{3}$ Theoretical Condensed Matter Physics and Computational Materials Physics Laboratory, College of Physical Sciences, University of Chinese Academy of Sciences, Beijing 100049, China.\\
$^{4}$ CAS Center for Excellence in Topological Quantum Computation, University of Chinese Academy of Sciences, Beijing 100190, China.\\
$^{5}$ Kavli Institute for Theoretical Sciences, University of Chinese Academy of Sciences, Beijing 100190, China.
}

\begin{abstract}
We theoretically investigate the temperature gradient-dependent or unidirectional Seebeck effect (USE) in a magnetic/ nonmagnetic topological insulator (TI) heterostructure with in-plane magnetization in  terms of  the semiclassical electron dynamics and Fermi golden rule. The USE has a quantum origin arising from the magnon asymmetric scattering of surface Dirac electrons on TI. We discuss the USE in the heterostructures, Cr$_{x}$(Bi$_{1-y}$Sb$_y$)$_{2-x}$Te$_{3}$/ (Bi$_{1-y}$Sb$_{y})_2$Te$_{3}$. The USE exhibits $\cos\phi$-dependence (measured from $y$-direction) on the orientation of magnetization. It's found that the sign of USE stays unchanged when the system is transferred from $p$-doping to $n$-doping. The USE shows on inverse-linearly temperature dependent at high temperature.
\end{abstract}

\maketitle

\section{Introduction}
\label{Introduction}
The three-dimensional (3D) topological insulators (TI)\cite{Hasan,Qi} represent a new class of 3D materials with an insulating bulk and conductive surface states. The surface Dirac electrons have their spin locked to their momentum, namely, spin-momentum locking [ Fig. ~\ref{2} (c)], hosting exotic topological quantum effects \cite{Fu1,Qi2,Yu,Yoshimi} and finding applications in spintronics and quantum computations. On the other hand, the interconversion of angular momentum between a conduction electron and local magnetization is one of the vital issues of contemporary research of spintronics research\cite{Wolf,Fert} and spin caloritronics \cite{Boona,G.E.W.Bauer,Xiao-Qin1,Xiao-Qin2}. Recently, the realization of a system, in which spin-polarized two dimensional (2D) Dirac electrons coexist with ferromagnetism on the surface of magnetic topological insulators (MTI) through doping magnetic impurities in TI \cite{Yu,C-Z-Change}, provides a new platform.

Owing to the interplay of the spin-momentum-locked surface states and magnetism, a series of novel transport phenomena are expected, such as quantum anomalous Hall effect (QAHE)\cite{C-Z-Change}, current-nonlinear Hall effect (CNHE) \cite{Yasuda1} and unidirectional magnetoresistance (UMR)\cite{Avci1,Avci2,Langenfeld,Yasuda2}. UMR has recently been proposed in TI heterostructures \cite{Fan,Yasuda1,Yasuda2,Yoshimi2,Yasuda3} composed of nonmagnetic TI (Bi$_{1-y}$Sb$_{y}$)$_{2}$Te$_{3}$ (BST) \cite{J-Zhang} and magnetic TI Cr$_{x}$(Bi$_{1-y}$Sb$_y$)$_{2-x}$Te$_{3}$ (CBST)\cite{C.Z.Chang2}, which describes the resistance-value dependence on the sign of the outer product of current $\mathbf{J}$ and the in-plane magnetization $\mathbf{M}$ vectors and is identified to originate from the asymmetric scattering of electron by magnons \cite{Yasuda2}. It is natural to ask whether the Seebeck effect depends on the relative orientations of the in-plane magnetization with respect to the temperature gradient in the heterostructures of BST/CBST.

In this paper, we theoretically investigate a temperature-gradient-direction dependent or unidirectional Seebeck effect (USE) induced by magnon asymmetric scattering in the heterostructure of BST/CBST. We believe
that the proposed effect is very useful in spin caloritronics.
We derive the formula of nonlinear longitudinal current $j_{x}$ response to the temperature gradient $\nabla_{x}T$ up to the second order based on the Boltzmann theory. We introduce, phenomenologically, a quantity $\Delta S$ which is expressed by the difference of Seebeck coefficients between the cases of the forward and backward temperature gradients $\Delta T$ [see Figs. \ref{USE}(a)\ref{USE}(b)], to characterize the  USE in Sec.~\ref{TD}.  We derive the formula of magnon relaxation time and determine the expressions of USE for BST/CBST heterostructure in Sec.~\ref{Model}. The behavior of USE is discussed in Sec.~\ref{RD}.

\section{Unidirectional Seebeck effect}
\label{TD}
With the relaxation time approximation, the Boltzmann equation for the distribution of electrons in absence of electric field reads ~\cite{Manhan,Xiao-Qin3}
\begin{equation}
f-f_{0}=-\tau(\mathbf{k}) \frac{\partial f}{\partial r_a}\cdot v_{a},
\label{Bol-eq3}
\end{equation}
where $\tau$ represents the relaxation time, and $v_{a}$ and $r_{a}$ denote the $a$ component of the velocity and coordinate position of electrons, respectively. $f_{0}$ is the local equilibrium distribution function. We are interested in the response up to the second order in temperature gradient, and hence have the nonequilibrium distribution function $f\approx f _{0}+f_{1}+f_{2}$ with the term $f_{n}$  to vanish as $(\partial{T}/\partial \mathbf{r}_{a})^{n}$. After a series of careful derivations in Appendix \ref{ndf}, the formulas of $f_{1}$ and $f_{2}$ can be determined and are given in Eq. (\ref{f1}).
The charge current $j_{a}$ in $a$ direction is
\begin{equation}
j_{a}=-e\int[d\mathbf{k}]v_{a}f(\mathbf{r},\mathbf{k}),
\label{J1}
\end{equation}
where $\int[d\mathbf{k}]$ is shorthand for $\int d\mathbf{k}/(2\pi)^{2}$. Based on Eqs. (\ref{J1}) and (\ref{f1}), the current $j_{x}$ in $x$ direction is found to be
\begin{equation}
j_{x}=\alpha_{xx}^{(1)}\left(-\nabla_{x}T\right)+\left(\alpha_{xx,1}^{(2)}+\alpha_{xx,2}^{(2)}\right)(\nabla_{x}T)^{2}-
\alpha_{xx,1}^{\left(2\right)}T\nabla^{2}_{x}T,
\label{cur-1}
\end{equation}
with
\begin{eqnarray}
\alpha_{xx}^{(1)}&=&e\int[d\mathbf{k}]\tau\left(\mathbf{k}\right)v_{x}\frac{\epsilon_\mathbf{k}-\mu}{\hbar T}\frac{\partial f_{0}}{\partial k_{x}},\nonumber\\
\alpha_{xx,1}^{(2)}&=&-e\int[d\mathbf{k}]\left(\tau\left(\mathbf{k}\right)\right)^{2}v^{2}_{x}\frac{\epsilon_\mathbf{k}-\mu}{\hbar T^{2}}\frac{\partial f_{0}}{\partial k_{x}},\nonumber\\
\alpha_{xx,2}^{(2)}&=&-e\int[d\mathbf{k}]\left(\tau\left(\mathbf{k}\right)\right)^{2}v_{x}\left(\frac{\epsilon_\mathbf{k}-\mu}{\hbar T}\right)^{2}\frac{\partial^{2} f_{0}}{\partial k_{x}^{2}}.\nonumber\\
\label{coeffient1}
\end{eqnarray}

Therefore, the relationship between temperature gradient and the thermoelectric voltage $\Delta V (=V_\text{right}-V_\text{left})$ can be expressed in a nonlinear form of
\begin{equation}
\begin{aligned}
\Delta V=\frac{l}{\sigma_{xx}}\left[\alpha_{xx}^{(1)}\left(-\nabla_{x}T\right)+\alpha_{xx}^{(2)}\left(\nabla_{x}T\right)^{2}\right].
\end{aligned}
\label{Delta-V}
\end{equation}

\begin{figure}
\centering
\includegraphics[width=1.0\linewidth,clip]{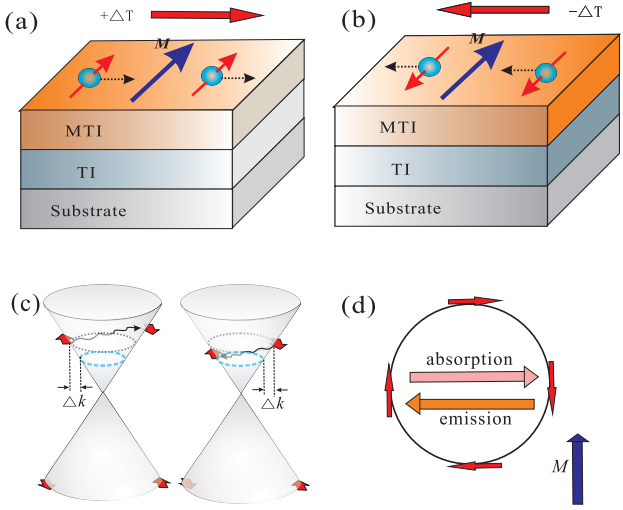}
\caption{Schematic illustration of the concept for USE in heterostructures topological insulator (TI)/ magnetic TI under (a) $+\Delta T$ and (b) $-\Delta T$  temperature gradient. Here, the magnetic field, magnetization, and temperature gradient are all aligned in plane. 
(c) Schematic illustration for the asymmetry magnon scattering of Dirac electrons on the spin-momentum-locked Fermi surface. (d) Top view of the scattering process for magnon emission and absorption.}
\label{USE}
\end{figure}

To obtain Eq. (\ref{Delta-V}), we have used $\Delta V=l E_{x}={j_{x}l}/{\sigma_{xx}}$ 
and meanwhile assume the uniform temperature gradient in the system, i.e., $\nabla^{2}_{x}T=0$. Here $\alpha_{xx}^{(2)}=\alpha_{xx,1}^{(2)}+\alpha_{xx,2}^{(2)}$ and $l$ is the length of the sample. Equation (\ref{Delta-V}) hints that the voltage drop is nonlinearly dependent on the temperature gradient, which gives USE: when modulating the direction of temperature gradient, the voltage absolute value generated through Seebeck effect is changed [Figs.~\ref{USE}(a) and (b)].  Therefore, the Seebeck coefficient $S=-\left({V_\text{left}-V_\text{right}}\right)/\left({T_\text{left}-T_\text{right}}\right),$ measured under plus or minus temperature difference $\Delta T$ [Figs. \ref{USE}(a) and (b)], is noticeably distinguishing and is temperature-gradient-direction dependent. Hence, the difference $\Delta S$ between the two temperature gradient directions can be applied to characterize the USE, which is determined by
\begin{equation}
\Delta S=S^{+}-S^{-}=\frac{2\alpha_{xx}^{(2)}\Delta T}{\sigma_{xx}l},
\label{DeltaS}
\end{equation}
with $S^{+}=\alpha_{xx}^{(1)}/\sigma_{xx}+\alpha_{xx}^{(2)}\Delta T/(l\sigma_{xx})$ and $S^{-}=\alpha_{xx}^{(1)}/\sigma_{xx}-\alpha_{xx}^{(2)}\Delta T/(l\sigma_{xx})$. Equation (\ref{DeltaS}) indicates that $\Delta S$ is linearly proportional to the temperature gradient. Here, we assume that the magnon scattering is completely independent of other scattering processes such as the impurity, phonon and so on. Thus, one could have
\begin{equation}
\frac{1}{\tau}=\frac{1}{\tau^{0}}+\frac{1}{\tau_{\text{mag}}},
\end{equation}
where $\tau^{0}$ is the nonmagnetic scattering relaxation time and $\tau_{\text{mag}}$ is the scattered relaxation time by magnons. The impurity scattering is considered to be dominant here, which gives $\tau^{0}<<\tau_\text{mag}$. To first-order approximation, therefore, the relaxation time can be written as $\tau=\tau^{0}-{(\tau^{(0)})^{2}}/{\tau_\text{mag}}$, leading to

\begin{widetext}
\begin{equation}
\begin{aligned}
\alpha^{(2)}_{xx}&=-e\left(\tau^{0}\right)^{2}\!\int [d\mathbf{k}]\left\{
\left[v_{x}^{2}\frac{\epsilon_\text{k}-\epsilon_{F}}{\hbar T^{2}}\frac{\partial f_{0}}{\partial k_{x}}+v_{x}\left(\frac{\epsilon_\text{k}-\epsilon_{F}}{\hbar T}\right)^{2}\frac{\partial^{2} f_{0}}{\partial k_{x}^{2}}\right]\!
-\!\frac{2\tau^{0}v_{x}}{\tau_\text{mag}}\!\left[v_{x}\frac{\epsilon_\text{k}-\epsilon_{F}}{\hbar T^{2}}\frac{\partial f_{0}}{\partial k_{x}}+\left(\frac{\epsilon_\text{k}-\epsilon_{F}}{\hbar T}\right)^{2}
\frac{\partial^{2} f_{0}}{\partial k_{x}^{2}}\!\right]\right\}.
\end{aligned}
\label{alph}
\end{equation}
\end{widetext}

Obviously, Eq. (\ref{alph}) involves two distinguished mechanisms that result in the nonlinear Seebeck effect and the USE:  the asymmetric energy dispersion along $k_x$ direction (the first term) and the asymmetric magnon scattering (the second term). In completely linear-$k$ Dirac dispersion, one can actually find that the first term in Eq. (\ref{alph}) is zero by exploiting the parity (Table \ref{parities}). Besides, even in the presence of in-plane magnetic field or magnetization, this term is still zero. This is because the Dirac point and the whole dispersion will shift in $k$-space simultaneously and consistently [Fig.~\ref{2}(c)] so that the velocity and the occupation are unchanged. In this work, we  consider only  the USE induced by the asymmetric magnon scattering, namely,
%

\begin{table}[tbph]
\centering
\caption{ Parity about $k_{x}$ for linear-k Dirac dispersion.
}%
\begin{tabular*}{8 cm}{@{\extracolsep{\fill}}lcccc}
 \hline\hline
 \text{function} &\text{parity}\\
 $\epsilon_{\mathbf{k}}$  &\text{even}\\
 $v_{x}$ &\text{odd}\\
 $\frac{\partial f_{0}}{\partial k_{x}}$ &\text{odd}\\
$\frac{\partial^{2} f_{0}}{\partial k_{x}^{2}}$ &\text{even}\\
 \hline\hline
 \end{tabular*}
 \label{parities}
 \end{table}

\begin{equation}
\begin{aligned}
\alpha^{(2)}_{xx,\text{mag}}&=2e\left(\tau^{0}\right)^{3}\int [d\mathbf{k}]\frac{1}{\tau_\text{mag}\left(\mathbf{k}\right)}\left[v^{2}_{x}\frac{\epsilon_\text{k}-\epsilon_{F}}{\hbar T^{2}}\frac{\partial f_{0}}{\partial k_{x}}\right.\\
&\left.+v_{x}\left(\frac{\epsilon_\text{k}-\epsilon_{F}}{\hbar T}\right)^{2}
\frac{\partial^{2} f_{0}}{\partial k_{x}^{2}}\right].
\end{aligned}
\label{alph-mag}
\end{equation}

\section{Model}
\label{Model}
\begin{figure}[h]
  \begin{minipage}[t]{0.42\linewidth}
 \centering
 \includegraphics[width=1.1\textwidth]{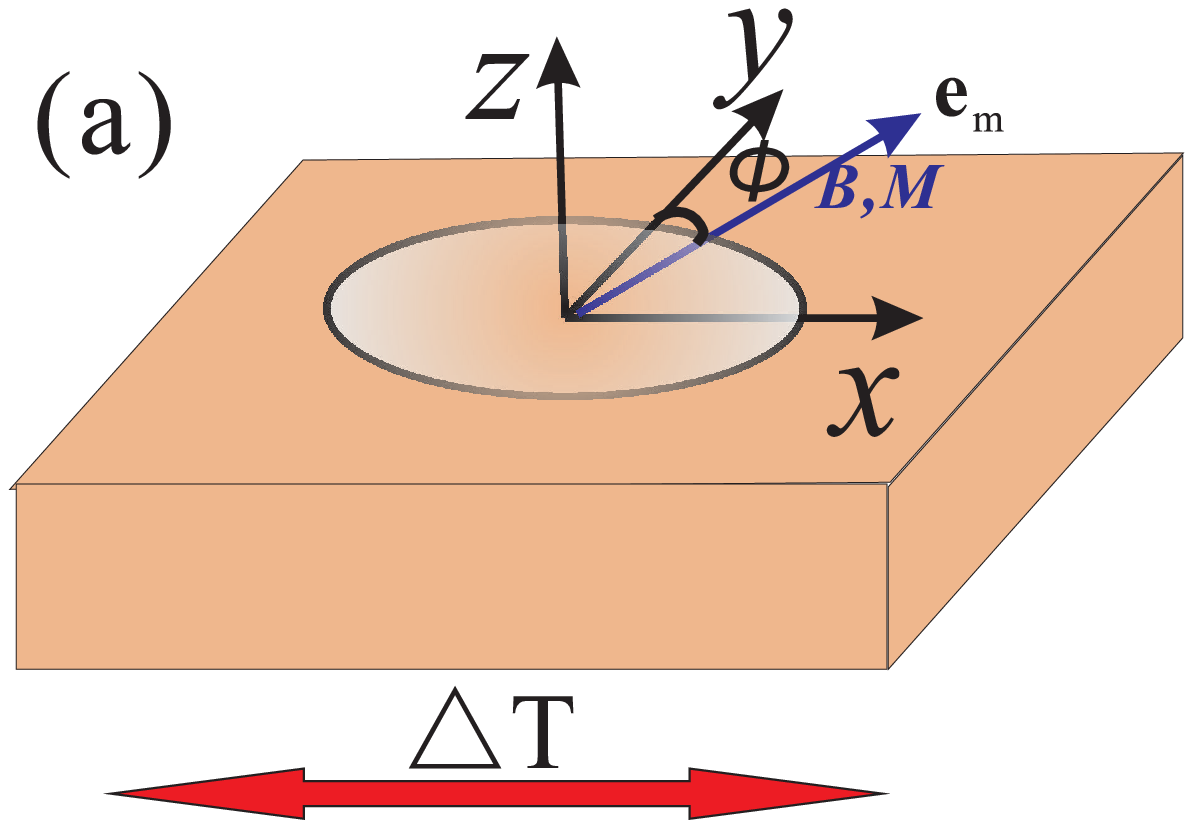}
 \end{minipage}
 \hfill
 \begin{minipage}[t]{0.5\linewidth}
 \centering
 \includegraphics[width=1.1\textwidth]{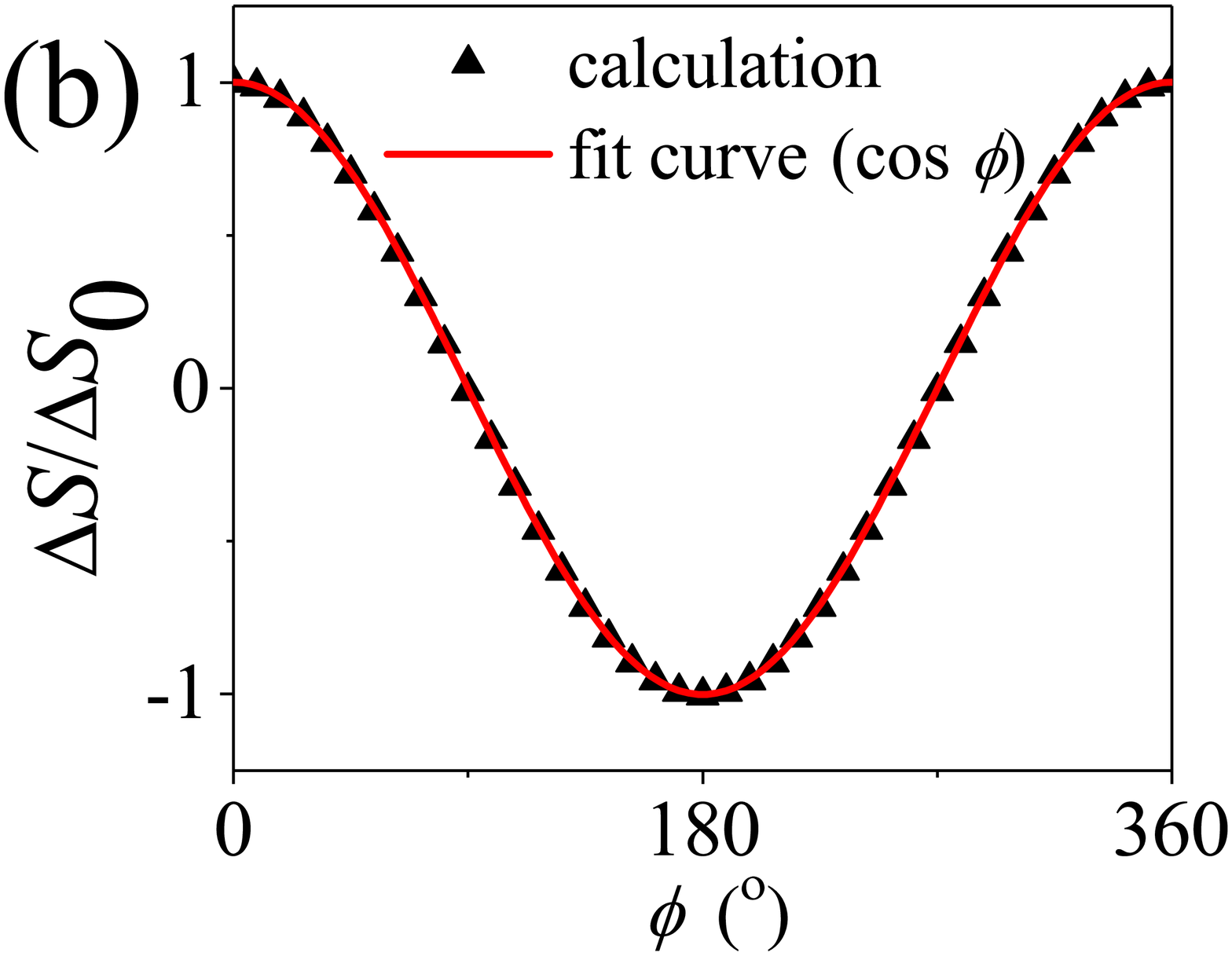}
 \end{minipage}
 \centering
 \includegraphics[width=0.56\textwidth]{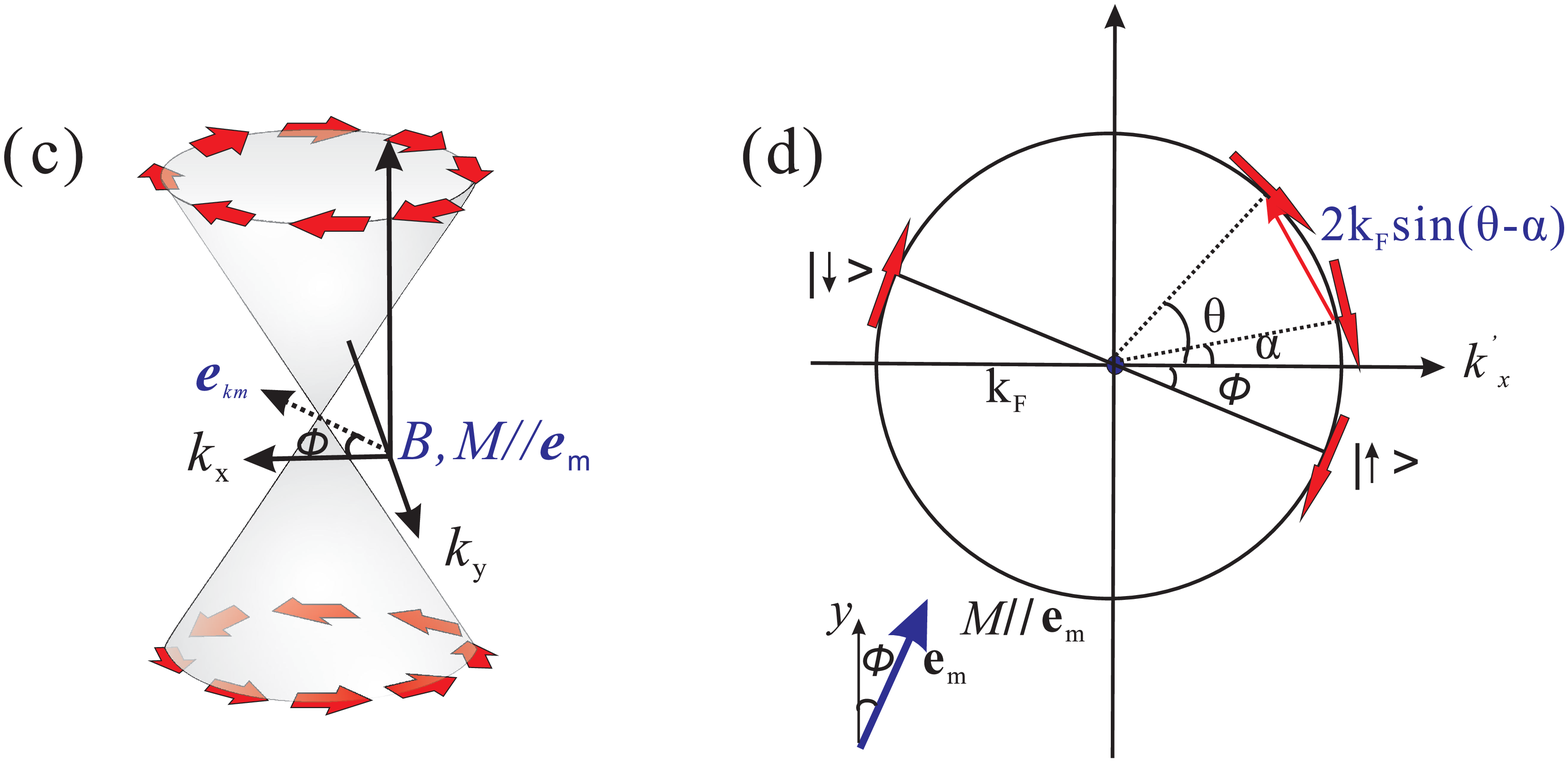}
\caption{(a) Schematic configuration for the measurement of in-plane magnetic field (or magnetization) $\phi$ dependence of $\Delta S$, the azimuth angle $\phi$ is measured from the $y$ axis. (b) The in-plane magnetization orientation (i.e., $\phi$) dependence of normalized $\Delta S/\Delta S_{0}$. Here  $\Delta S_{0}$ is $\Delta S$ at $\phi=0$. (c) Schematic diagram of the band structure under in-plane magnetization $M//\mathbf{e}_{m}=(\sin\phi,\cos\phi)$ and spin-momentum locking of the surface Dirac state in TI. (d) Top view of the magnon scattering in 2D Dirac dispersion. The orientation of spin eigenfunctions $\mid\uparrow>$ and $\mid\downarrow>$ are illustrated when magnetization is parallel to $\mathbf{e}_{m}$.}
\label{2}
\end{figure}

One of candidate materials to observe 
the USE originating from the asymmetric scattering of electrons by magnons is the TI heterostructure BST/CBST. 
By tuning the composition $y$, one can modulate the Fermi energy $E_{f}$ of the surface state inside the bulk band gap \cite{Yasuda1,Yasuda2,Yoshimi2,Yasuda3,J-Zhang}. Hence, the carriers from the top and bottom surface states with single Dirac cone will dominate the conduction. In addition, owing to the heterostructure, only one surface involved in MTI layer interacts effectively with magnetic interaction \cite{Yoshimi2,Yasuda3}. Besides, the magnetization $\mathbf{M}$ of CBST 
initially points along the $z$ direction and leads to an exchange gap in the surface Dirac state. When the in-plane magnetic field $B$ is applied up to $B_{0}$ ($\sim0.7$ T) \cite{Yasuda2}, the orientation of magnetization will gradually be changed into the in-plane direction and the Dirac surface states will eventually become gapless. In this work, we consider the situation in which the magnetization is already oriented in plane, with $\mathbf{M}=(m_{x},m_{y})=m\vec{e}_\text{m}$ and $\vec{e}_\text{m}=(\sin\phi,\cos\phi)$ [see Fig.~\ref{2} (a)], where the azimuth angle $\phi$ is measured from $y$-axis. Owing to the coupling between the electron spin and localized magnetic moments, the Hamiltonian of the Dirac surface electrons is affected by the magnetization $\mathbf{M}$ and can be written as
\begin{equation}
H_{0}=\left(m\sin\phi+v_{F}\hbar k_{y}\right)\hat{\sigma}_{x}+\left(m\cos \phi-v_{F}\hbar k_{x}\right)\hat{\sigma}_{y},
\end{equation}
where $v_{F}$ is the Fermi velocity, $\hbar$ represents the Planck constant, $\hat{\sigma}$ denotes the Pauli matrices for the two basis functions of the energy band, $m$ indicates the magnitude of magnetization $\mathbf{M}$, and the definition of the azimuthal angle $\phi$ is given in Fig. \ref{2}(a). For simplicity, we also ignore $k^{2}$ term and hexagonal warping ($k^{3}$ term) in the surface state of BST/CBST. The energy eigenvalues are
\begin{equation}
\epsilon_{\mathbf{k}}=n\sqrt{\left(v_{F}\hbar k_{x}-m\cos\phi\right)^{2}+(v_{F}\hbar k_{y}+m\sin\phi)^{2}},
\label{E-eig}
\end{equation}
where $n\left(=\pm1\right)$ is the band index. Equation (\ref{E-eig}) hints that the Dirac point and the whole dispersion shifts towards the $\vec{e}_{k_{m}}=(\cos\phi,-\sin\phi)$ direction with the magnetic field or magnetization along direction $\vec{e}_\text{m}$ [Figs.\ref{2}(a) and (c)]. In the following, we consider the interaction between the surface conduction electron and the localized spin composed of Cr $d$ orbits. When \textbf{M} is along the $\vec{e}_\text{m}$ direction, the localized spin is pointing in the $-\vec{e}_\text{m}$ direction. Therefore, the angular momentum of the magnon is $+1$. Owing to the conservation of angular momentum, the interaction Hamiltonian $H^{'}$ is
\begin{equation}
\begin{aligned}
H^{'}&\approx\sum_{i}j_\text{ex}\left(c^{+}_{i,\uparrow}c_{i,\downarrow}b_{i}+c^{+}_{i,\downarrow}c_{i,\uparrow}b^{+}_{i}\right)\\
&=\sum_{\mathbf{k}\mathbf{q}}j_\text{ex}\left(c^{+}_{\mathbf{k}+\mathbf{q},\uparrow}c_{\mathbf{k},\downarrow}b_\mathbf{q}
+c^{+}_{\mathbf{k-q},\downarrow}c_{\mathbf{k},\uparrow}b^{+}_\mathbf{q}\right),
\end{aligned}
\label{H1}
\end{equation}
where $j_\text{ex}$ is the exchange coupling constant, and $b^{+}\left(b\right)$ and $c^{+}\left(c\right)$ denote the creation (annihilation) operator of the magnon and surface Dirac electron, respectively. Equation (\ref{H1}) involves two processes: magnon absorption ($c^{+}_{\mathbf{k}+\mathbf{q},\uparrow}c_{\mathbf{k},\downarrow}b_\mathbf{q}$) and magnon emission ($c^{+}_{\mathbf{k-q},\downarrow}c_{\mathbf{k},\uparrow}b^{+}_\mathbf{q}$). In the magnon absorption process, when the Dirac electron from the surface states $|\mathbf{k},\downarrow>$ is scattered to state $|\mathbf{k}+\mathbf{q},\uparrow>$, a magnon with momentum $\mathbf{q}$ is absorbed due to the conservation of momentum and angular momentum [left plane of Fig. ~\ref{USE} (c)]. In the magnon emission process, the electron spin is reversed from $\uparrow (s_{\phi}=1/2)$ to $\downarrow (s_{\phi}=-1/2)$ by the emission of magnon [right plane of Fig. ~\ref{USE} (c)] resulting from the spin-momentum locking of surface Dirac state and the conversation of angular momentum.

Based on the Fermi golden rule, the magnon relaxation time $\tau_\text{mag}$ is found to be \cite{Yasuda1,Yasuda2}
\begin{equation}
\frac{1}{\tau_\text{mag}\left(\mathbf{k}\right)}\approx\sum_{\mathbf{k}^{\prime}}W_\text{mag}\left(\mathbf{k}^{\prime}|\mathbf{k}\right)
\left[1-f(\mathbf{k}^{\prime})\right],
\label{tau}
\end{equation}
with
\begin{equation}
\begin{aligned}
W_\text{mag}\left(\mathbf{k}^{\prime}|\mathbf{k}\right)&=W_\text{abs}(\mathbf{k}^{\prime},\sigma^{\prime};
n_{\mathbf{k}^{\prime}-\mathbf{k}}-1\mid \mathbf{k},\sigma;n_{\mathbf{k}^{\prime}-\mathbf{k}})\\
&+W_\text{emit}(\mathbf{k}^{\prime},\sigma^{\prime};
n_{\mathbf{k}-\mathbf{k}^{\prime}}+1\mid \mathbf{k},\sigma;n_{\mathbf{k}-\mathbf{k}^{\prime}}),
\end{aligned}
\label{Wm}
\end{equation}
where $1-f(\mathbf{k}^{\prime})$ represents the probability of final state of the electron in which the electron is unoccupied. $n_{\mathbf{k}}$ denotes the number of magnon, $W_\text{abs} (W_\text{emit})$ are the scattering probability for the magnon absorption (emission) process, respectively, and are characterized as
\begin{equation}
\begin{aligned}
W_\text{abs}&=\frac{2\pi}{\hbar}|\langle\mathbf{k}^{\prime},\sigma^{\prime};
n_{\mathbf{k}^{\prime}-\mathbf{k}}-1|H^{\prime}|\mathbf{k},\sigma;n_{\mathbf{k}^{\prime}-\mathbf{k}}\rangle|^{2}\\
&\times\delta(\epsilon_{\mathbf{k}^{\prime}}-\epsilon_\mathbf{k}-\hbar\omega_{\mathbf{k}^{\prime}-\mathbf{k}}),\\
W_\text{emit}&=\frac{2\pi}{\hbar}|\langle\mathbf{k}^{\prime},\sigma^{\prime};
n_{\mathbf{k}-\mathbf{k}^{\prime}}+1|H^{\prime}|\mathbf{k},\sigma;n_{\mathbf{k}-\mathbf{k}^{\prime}}\rangle|^{2}\\
&\times\delta(\epsilon_{\mathbf{k}^{\prime}}-\epsilon_\mathbf{k}+\hbar\omega_{\mathbf{k}-\mathbf{k}^{\prime}}).
\end{aligned}
\label{W-AE}
\end{equation}
Taking the interaction Hamiltonian $H^{\prime}$ (Eq. (\ref{H1})) into Eq. (\ref{W-AE}), we have
\begin{equation}
\begin{aligned}
W_\text{abs}&\!=\!\frac{2\pi}{\hbar}\!j^{2}_{\text{ex}}n_{\mathbf{k}^{\prime}-\mathbf{k}}|\langle\sigma^{\prime}\mid c^{+}_{\uparrow}c_{\downarrow}\mid \sigma\rangle\mid^{2}\delta\left(\epsilon_{\mathbf{k}^{\prime}}\!-\!\epsilon_\mathbf{k}\!-\!\hbar\omega_{\mathbf{k}^{\prime}-\mathbf{k}}\right), \\
W_\text{emit}&=\frac{2\pi}{\hbar}j^{2}_{\text{ex}}\left(n_{\mathbf{k}-\mathbf{k}^{\prime}}+1\right)|\langle\sigma^{\prime}\mid c^{+}_{\downarrow}c_{\uparrow}\mid \sigma\rangle\mid^{2}\\
&\times\delta\left(\epsilon_{\mathbf{k}^{\prime}}-\epsilon_\mathbf{k}-\hbar\omega_{\mathbf{k}-\mathbf{k}^{\prime}}\right),
\end{aligned}
\label{Wae}
\end{equation}
where $\mid\sigma\rangle $  ( $\mid\sigma^{\prime}\rangle$ ) indicates the initial (final) electron spin state, respectively. The spin orientation of surface electron in MTI around the Dirac cone is not fixed but rotates around the $z$ axis [see Fig. \ref{2}(c)]. In the following, we consider the scattering from the position $\alpha$ to $\theta$ as shown in Fig.~\ref{2} (d), namely $\mid\sigma\rangle=\mid\alpha\rangle$ and $\mid\sigma^{\prime}\rangle=\mid\theta\rangle$. Here, the spin eigenfunction at $\alpha$ (or $\theta$) in $\{\mid\uparrow\rangle,\mid\downarrow\rangle\}$ representation is
\begin{equation}
\begin{aligned}
\mid\alpha\rangle&=\sin\frac{\phi+\alpha}{2}\mid\uparrow\rangle+\cos\frac{\phi+\alpha}{2}\mid\downarrow\rangle,\\
\mid\theta\rangle&=\sin\frac{\phi+\theta}{2}\mid\uparrow\rangle+\cos\frac{\phi+\theta}{2}\mid\downarrow\rangle,
\end{aligned}
\label{alpha}
\end{equation}
where $\mid\uparrow\rangle$ ($\mid\downarrow\rangle$) represents the state in which the spin direction is antiparallel (parallel) to the in-plane magnetic field, respectively. Therefore, taking the formulas of $\mid\sigma\rangle=\mid\alpha\rangle$ ($\mid\sigma^{\prime}\rangle=\mid\theta\rangle$) determined by Eq.(\ref{alpha}) into Eq. (\ref{Wae}) and accompanyied with a series of derivation in Appendix \ref{MRT}, the magnon relaxation time in Eq. (\ref{tau}) is found to be
\begin{equation}
\frac{1}{\tau_\text{mag}\left(\alpha,\phi,\Delta k\right)}=\frac{1}{\tau^\text{m}_{F}}\int^{2\pi}_{0} d\theta V_\text{mag}\left(\theta+\phi,\alpha+\phi,\Delta k\right),
\label{tau1}
\end{equation}
with $\frac{1}{\tau^\text{m}_{F}}=\frac{k_{F}}{\left(2\pi\right)}\frac{j^{2}_\text{ex}A}{v_{F}\hbar^{2}}$, $A$ is the area of the sample, and the integrand $V_\text{mag}\left(\theta,\alpha,\Delta k\right)$ is given in Eq. (\ref{APP-B-V}).

\section{Results and Discussion}
\label{RD}
 By approximating the conductivity as $\sigma_{xx}\approx(e^{2}/4\pi\hbar) v_{F}k_{F}\tau^{0}$ [see Eq. (\ref{APP-B-Con})], we found the quantity $\Delta S$ determining the USE in Eq. (\ref{DeltaS}) to be (see details in Appendix \ref{expresion-USE})

\begin{equation}
\begin{aligned}
\frac{\Delta S}{\Delta T}&=-\zeta\frac{A \epsilon_{F}}{lT}
\!\int\!\!\int\!\int dx d\alpha d\theta V_\text{mag}\left(\theta+\phi,\alpha+\phi,\frac{k_\text{B}Tx}{\hbar v_{F}}\right)\\
&\times\frac{xe^{x}\cos^{3}\alpha }{\left(e^{x}+1\right)^2} \left(1+
\frac{1-e^{x}}{e^{x}+1}x\right),
\end{aligned}
\label{DeS-F}
\end{equation}
where $x=\hbar v_{F}\Delta k/(k_\text{B}T)$ with $\Delta k$ measured from $k_{F}$ (Fermi wavenumber). The typical scale $\zeta=4k_\text{B}j^{2}_\text{ex} \left(\tau^{0}\right)^{2}/(e\hbar^{3}v_{F})$  is around  1.2$\times 10^{9}[\text{eKm}]^{-1}$ for heterostructures BST/CBST. We use the following typical values: the Fermi velocity $v_{F}\approx 1\times 10^{5}\text{m}/\text{s}$ \cite{Yoshimi3}, and the nonmagnetic scattering relaxation time $\tau^{0}\approx 10^{-13} s$ estimated by $\tau^{0}=\mu m/e$. Mobility $\mu$ of (Bi$_{1-x}$Sb$_{x}$)$_{2}$Te$_{3}$ ranges from 100 to 500 cm$^2$V$^{-1}$s$^{-1}$ ~\cite{J-Zhang} when tuning the composition $x$. We use $\mu=300$ cm$^2$V$^{-1}$s$^{-1}$ for an estimation. 
 Although the value of the exchange-coupling energy $j_{\text{ex}}$ is not well known for the heterostructures BST/CBST. Here, we adopt $j_{\text{ex}}\approx0.1$ eV from Ref. [\onlinecite{Qin-Liu}] which is a typical value for exchange-coupling of surface states Sb$_2$Te$_3$ and the magnetic impurities.

To disclose the microscopic origin of USE induced by magnon scattering intuitively, we firstly neglects the magnon dispersion and use $g \mu_{B}B$ as a magnon energy \cite{Skrotskii} for simplicity. Thus, ${\Delta S}/{\Delta T}$ in Eq. (\ref{DeS-F}) can be further simplified  as ( see details in Appendix \ref{expresion-USE})
%

\begin{equation}
\begin{aligned}
\frac{\Delta S}{\Delta T}&=-\zeta\frac{3\pi A \tau_{F}^{m}\epsilon_{F}}{8lT}\cos\phi\int dx (\frac{1}{\tau^{+}}-\frac{1}{\tau^{-}})\frac{xe^{x}}{\left(e^{x}+1\right)^2}\\
&\times\left(1+
\frac{1-e^{x}}{e^{x}+1}x\right),
\end{aligned}
\label{non-Se}
\end{equation}
with
\begin{eqnarray}
\frac{1}{\tau^{+}}&=&\frac{1}{e^{\beta\hbar\omega}-1}\left(1-
\frac{1}{e^{\left(x+\beta\hbar\omega\right)}+1}\right)\frac{1}{\tau_{F}^{m}},\label{taup}\\
\frac{1}{\tau^{-}}&=&\!\left(\frac{1}{e^{\beta\hbar\omega}-1}+1\!\right)\!\left(1-
\frac{1}{e^{\left(x-\beta\hbar\omega\right)}+1}\!\right)\frac{1}{\tau_{F}^{m}}\label{taum}.
\end{eqnarray}
Here, $\tau^{+} \left(\tau^{-}\right)$ is the relaxation time of magnon scattering from the left (right) branch to the right (left) one [Fig. \ref{USE}(d)]. The first factors of Eqs. (\ref{taup}) and (\ref{taum}) give the probability of magnon absorption and emission, respectively, and the second ones show the probability that the final state of the electrons is unoccupied.

Since $1/\tau^{+}$ and $1/\tau^{-}$ are not equal in general, Equation (\ref{non-Se}) gives finite USE which attributes to the magnon asymmetric scattering of electron in the TI, namely the scattering rates are different in magnon absorption and emission processes.  Eq. (\ref{non-Se}) also hints the in-plane magnetization directional dependence of normalized $\Delta S/\Delta S_{0}\propto \cos\phi$,  $\Delta S_{0}$ is $\Delta S$ at $\phi=0^{\circ}$. 
Thus, when the magnetization $M$ is parallel or antiparallel to the $y$ axis (i.e., $\phi=0\,\,,\pi\,,2\pi$), the $|\Delta S|$ will reach its maximum. However, the USE will disappear when magnetization is aligned to the direction of temperature gradient (i.e., $M//x$). The $\cos\phi$-dependence on the orientation of magnetization of the USE can be ascribed to the asymmetric magnon scattering  induced by the $m_{y}(\propto\cos\phi)$ part of magnetization. The magnon scattering rate $1/\tau_\text{mag}$ can be divided into two parts: symmetric part $1/\tau^\text{S}_\text{mag}$ [Eq. (\ref{APP-D-AS})] and antisymmetric part $1/\tau^\text{A}_\text{mag}$ [Eq. (\ref{APP-D-AS})] when reversing $k_{x}$ (a consequence of a mirror operation with respect to $k_{y}-k_{z}$ plane in momentum space). $m_{x}(\propto\sin\phi)$ ($m_{y}$)of magnetization only contributes to the symmetric (antisymmetric) scattering. Thus, only the antisymmetric part makes contribution to the USE [see details in Appendix \ref{symmetry-cos} ]. Therefore, the USE shows $\cos\phi$-dependence on the orientation of magnetization.

\begin{figure}[tb]
\centering
\includegraphics[width=1.0\linewidth,clip]{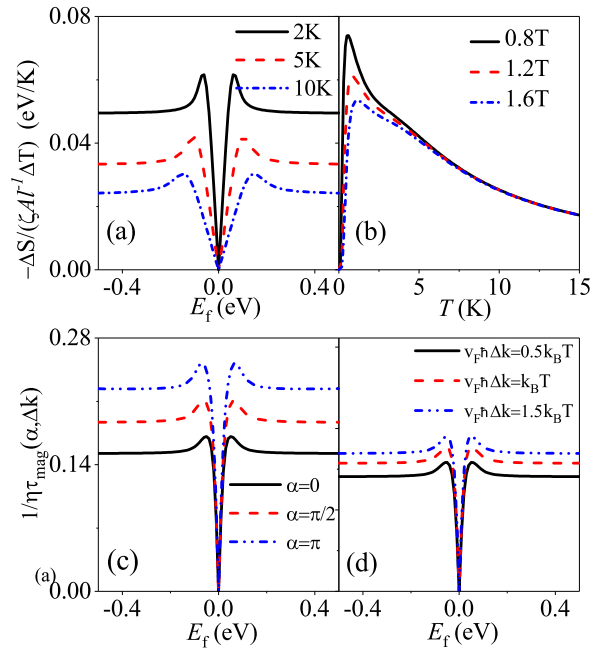}
\caption{(a) $\Delta S/ \Delta T$ as a function of the Fermi energy for different temperature with the magnetic field $B=1T$. (b) $\Delta S/ \Delta T$ as a function of temperature gradient for different magnetic field. The magnetic field in (a) is taken at $1T$. The Fermi energy $E_{f}$ is fixed at $0.1 eV$ in (b). Magnon scattering rate $1/\eta\tau_\text{mag}(\alpha,\Delta k)$ as function of Fermi energy for different polar angle $\alpha$ in (c) and $\Delta k$ in (d). $\alpha$ is polar angle measured from $k_{x}$ axis and $\Delta k$ is the radium measured from Fermi momentum $k_{F}$. Where $\eta=\epsilon_{0}j^{2}_\text{ex}A/(2\pi v_{F}^{2}\hbar^{3})$ with $\epsilon_{0}=1eV$ in (c)(d). The energy $v_{F}\hbar \Delta k$ is fixed at $1.5 k_\text{B}T$ in (c) and $\alpha=0$ in (d). }
\label{SE}
\end{figure}

In the following, we investigate the USE without neglecting the magnon band dispersion. In the long wavenumber case, the band dispersion of magnon is \cite{Onose}
\begin{equation}
\begin{aligned}
\hbar\omega&=D_{s}q^{2}+g\mu_\text{B}B\\
&=4D_{s}k^{2}_{F}\sin^{2}\left(\theta-\alpha\right)+g\mu_\text{B}B,
\end{aligned}
\end{equation}
where $D_{s}$ is the spin stiffness constant, $g\approx2$ for the localized Cr moment in CBST. Figure \ref{2} shows the in-plane magnetization orientation dependence of normalized $\Delta S$ ($\propto \cos\phi$) when considering band dispersion of magnon. The result is consistent with the case in which the magnon dispersion is neglected. When the magnetic field or magnetization is aligned along the $y$ direction, the signal of $\Delta S$ reaches the maximum.

Figure \ref{SE}(a) shows the variation of $\Delta S$ with Fermi energy at different temperatures. The sign of $\Delta S$ does not change when the system is transferred from electron ($E_{f}<0$) into hole ($E_{f}>0$). This can be understood by considering the scattering process for holes in the same way as shown in Fig. \ref{USE}(d) 
for electrons. With modulating the Fermi energy through gate voltage to appropriate value (close to Dirac point), the signal of unidirectional Seebeck effect $\Delta S$ can reach its maximum for both electrons and holes. The appearance of peak in Fig. \ref{SE} (a) can be qualitatively understood as follow: with increasing the absolute value of Fermi $|E_{f}|$, the energy of magnon contributing to the scattering processes increases so that the related magnon population decrease and the density of state of carries (holes or electrons), on the contrary, increases. The combination of these two mechanisms leads to that the magnon scattering rate $1/\tau_\text{mag}$ [Fig. \ref{SE} (c)] increases rapidly firstly, then gradually decreases due to the increase of magnon population, giving rise to a peak feature of $\Delta S/\Delta T$.

The temperature $T$ dependence of $\Delta S$ at different magnetic fields is shown in Fig. \ref{SE}(b). As expected, the unidirectional Seebeck effect tends to zero when $T$ approaches zero owing to frozen magnon. Indeed, in the extremely low temperature regime $T\ll\hbar\omega_\text{min}$ ($\hbar\omega_\text{max}=g\mu_\text{B}B$), namely $T\ll 1.34$ K for $B=1$T,  $\Delta S$ tends to zero owing to the frozen magnon. At the limit of high temperature $T\gg\hbar\omega_\text{max}$ ($\hbar\omega_\text{max}=2D_{s}k^{2}_{F}+g\mu_\text{B}B$ is the maximum energy of magnon), $\Delta S$, as expected, varies inversely proportional with $T$. Therefore, a  peak of $\Delta S$ will develop at finite temperature. And the peak shifts to higher temperature with increasing the magnetic field.
To understand the temperature dependence behaviours of the USE in high temperature qualitatively, we go back to the case in which the magnon dispersion is ignored for a more transparent picture. In the considered temperature regime, the number of magnons linearly depends on temperature ($n_{B}\approx k_{B}T/\hbar \omega$) and the difference between $1/(\exp(x+\beta\omega\hbar)+1)$ and $1/(\exp(x-\beta\omega\hbar)+1)$ is inversely dependent on T, leading to the temperature independence of $1/\tau^{+}-1/\tau^{-}$in Eq. (\ref{non-Se}), giving rise to the inverse-linear temperature dependence of $\Delta S$. Besides, the impact of varying magnetic field is also insignificant in this temperature regime. 

To numerically estimate the proposed effect, we take $\Delta S/\zeta A l^{-1}\Delta T \approx 0.3 \text{eV/K}$ (Fig. \ref{SE} (a)) for $T=10$ K. We also note that the USE in heterostructures BST/CBST can be estimated by the difference of voltage $V_\text{USE}$ before and after reversing the direction of temperature gradient as follows: $V_\text{USE}=|\Delta S \Delta T|$. The typical $\zeta$ in heterostructures BCT/CBST has been estimated to be order of  $1.2\times 10^{9}$ [\text{eKm}]$^{-1}$. Thus $V_\text{USE}\simeq$ 72 mV  with $\Delta T=10$ \text{mK} and w$=A/l=0.2 \mu\text{m}$ being the width of sample, which is measurable \cite{K. Uchida}.

\section{Conclusion}
In summary, we study the unidirectional Seebeck effect in the heterostructures of TI/MTI. It is found that the Seebeck coefficient $S$ is the temperature-gradient-direction dependent and has noticeable distinguishing feature when measured under positive ($+x$ axis) or negative (-$x$ axis) temperature difference $\Delta T$. We have derived this difference $\Delta S=S^{+}-S^{-}$ to characterize the unidirectional Seebeck effect induced by the magnon asymmetric scattering through the semi-classical framework of electron dynamics. Moreover, the quantity $\Delta S=S^{+}-S^{-}$ is strongly dependent on the orientation of the in-plane magnetization. When the in-plane magnetic field or magnetization is vertical to the temperature gradient, $|\Delta S|$ reaches its maximum. However, the signal of $\Delta S$ will disappear when applying an magnetic field  collinear to the temperature gradient. Fixing the magnetic field to $y$ direction, the unidirectional Seebeck effect is inverse-linearly dependent on temperature and insensitive to the magnetic field in the ``high" temperature regime ($T\gg\hbar \omega_\text{max}$) in which $T$ is far larger than the maximum energy of magnon contributing to scattering.

\section{acknowledgements}
This work is supported by the Fundamental Research Funds for the Central Universities and the NSFC (Grants No. 11674317 and No. 11974348). GS and ZGZ are supported in part by the National Key R\&D Program of China (Grant No. 2018FYA0305800), the Strategic Priority Research Program of CAS (Grant No. XDB28000000),  the NSFC (Grant No. 11834014), and the Beijing Municipal Science and Technology Commission (Grant No. Z118100004218001).

\appendix

\setcounter{equation}{0}
\setcounter{figure}{0}
\setcounter{table}{0}
\makeatletter
\renewcommand{\theequation}{A\arabic{equation}}
\renewcommand{\thefigure}{A\arabic{figure}}
\renewcommand{\thetable}{A\arabic{table}}

\bigskip
\bigskip

\noindent
\section{The non-equilibrium distribution function in the presence of temperature gradient} \label{ndf}
With the relaxation time approximation, the Boltzman equation for the distribution of electrons in absence of the electric field is
\begin{equation}
f-f_{0}=-\tau \frac{\partial f}{\partial r_{a}}\cdot v_{a}.
\label{Bol-eq3}
\end{equation}
To the response up to the second order in temperature gradient $\nabla T$, the local distribution function $f\left(\mathbf{r},\mathbf{k}\right)$ is written as
\begin{equation}
\begin{aligned}
f\left(\mathbf{k},\mathbf{r}\right)&=f_{0}\left(\mathbf{k},\mathbf{r}\right)+A_{a}\frac{\partial T}{\partial r_{a}}+
B_{a {\color{blue}{b}}}\frac{\partial{T}}{\partial{r_{a}}}\frac{\partial{T}}{\partial{r_{b}}}+O[(\partial_{a}T)^{3}]\\
&=f_{0}\left(\mathbf{k},\mathbf{r}\right)\!+\!f_{1}\left(\partial_{a}T\right)\!+\!f_{2}\left(\partial_{a}T\partial_{b}T\right)
+O[(\partial_{a}T)^{3}],
\end{aligned}
\label{non-equ-f-1}
\end{equation}
with
\begin{equation}
\left\{
\begin{aligned}
f_{1}\left(\partial_{a}T\right)&=A_{a}\partial_{a}T,\\
f_{2}\left(\partial_{a}T\partial_{b}T\right)&=B_{a b}\partial_{a}T\partial_{b}T,\\
\partial_{a}&\rightarrow \frac{\partial}{\partial r_{a}}.
\end{aligned}
\right.
\end{equation}
where $f_{0}\left(\mathbf{k},\mathbf{r}\right)$ is the local equilibrium distribution function, which is itself fixed by the temperature at $\mathbf{r}$ \cite{Ziman}, giving rise to
\begin{equation}
\frac{\partial f_{0}}{\partial r_{a}}=\frac{\partial f_{0}}{\partial T}\frac{\partial T}{\partial r_{a}}
\end{equation}
with
\begin{equation}
\frac{\partial f_{0}}{\partial T}=\frac{\left(\epsilon_\mathbf{k}-\mu\right)}{k_\text{B}T^{2}}(1-f_{0})f_{0}=-\frac{\left(\epsilon_\mathbf{k}-\mu\right)}{T}\frac{\partial f_{0}}{\partial \epsilon_\mathbf{k}}.
\label{Partial-I}
\end{equation}
Substituting the formula of $f$ in Eq. (\ref{non-equ-f-1}) into Eq. (\ref{Bol-eq3}) and comparing the expansion coefficients in the first-order of $\partial_{a} T$, one obtains
\begin{equation}
\begin{aligned}
f_{1}\left(\partial_{a}T\right)
=&-\tau\frac{\partial f_{0}}{\partial r_{a}}\cdot v_{a}+O[\partial_{a} T\partial_{b}T].\\
\end{aligned}
\end{equation}
Thus, we can have
\begin{equation}
f_{1}\left(\partial_{a}T\right)=-\tau\frac{\partial f_{0}}{\partial T}\partial_{a}T\cdot v_{a}.
\label{non-equ-f1}
\end{equation}
By iteration, then, we can have
\begin{equation}
\begin{aligned}
f_{2}\left(\partial_{a}T\partial_{b}T\right)&=-\tau\frac{\partial f_{1}}{\partial r_{a}}\cdot v_{a}\\
&={\tau^{2}}\left(\frac{\partial^{2}f_{0}}
{\partial T^{2}}{\partial_{a}T}{\partial_{b}T}+\frac{\partial f_{0}}{\partial T}\partial_{ab}T\right)v_{b}v_{a}.
\end{aligned}
\label{f22}
\end{equation}
Here, we introduce a trick to transfer ${\partial f_{0}}/{\partial T}$ into ${\partial f_{0}}/{\partial \mathbf{k}}$ 
through a partial differential treatment£º
\begin{equation}
\begin{aligned}
\frac{\partial f_{0}}{\partial \mathbf{k}}=\frac{\partial f_{0}}{\partial \epsilon_\mathbf{k}}\cdot \frac{\partial \epsilon_\mathbf{k}}{\partial \mathbf{k}}
=-\frac{\partial f_{0}}{\partial T}\frac{\hbar\mathbf{v}T}{\left(\epsilon_\mathbf{k}-\mu\right)}.
\end{aligned}
\label{IK-tran}
\end{equation}
In the above, we have used the relation: $\frac{\partial f_{0}}{\partial T}=-\frac{\left(\epsilon_\mathbf{k}-\mu\right)}{T}\frac{\partial f_{0}}{\partial \epsilon_\mathbf{k}}$ [Eq. \ref{Partial-I}] and $\frac{\partial \epsilon_\mathbf{k}}{\partial \mathbf{k}}={\hbar \mathbf{v}}$.
From Eq. (\ref{IK-tran}), it is easily to obtain the following identities:
\begin{equation}
\begin{aligned}
\frac{\partial{f_{0}}}{\partial T} \cdot v_{a}&=-\frac{\epsilon_\mathbf{k}-\mu}{\hbar T}\frac{\partial f_{0}}{\partial k_{a}},\\
\frac{\partial^{2}f_{0}}{\partial T^{2}}v_{a}v_{b}&=\frac{\epsilon_\mathbf{k}-\mu}{\hbar T^{2}}\frac{\partial f_{0}}{\partial k_{a}}v_{b}+\left(\frac{\epsilon_\mathbf{k}-\mu}{\hbar T}\right)^{2}\frac{\partial^{2} f_{0}}{\partial k_{a}\partial k_{b}}.
\end{aligned}
\end{equation}
Taking these identities into the formulas of $f_{1}$ [Eq. (\ref{non-equ-f1})] and $f_{2}$ [Eq. (\ref{f22})], one obtains
\begin{equation}
\begin{aligned}
f_{1}=&\frac{\tau}{T\hbar}\left(\epsilon_\mathbf{k}-\mu\right)\frac{\partial f_{0}}{\partial k_{a}} \partial_{a} T,\\
f_{2}=&-\frac{\tau^{2}}{T\hbar}\left(\epsilon_\mathbf{k}-\mu\right)v_{b}\frac{\partial f_{0}}{\partial k_{a}}\left(\partial_{ab}T-\frac{1}{T}\partial_{a}T\partial_{b}T\right)\\
&+\frac{\tau^{2}}{\hbar^{2} T^{2}}\left(\epsilon_\mathbf{k}-\mu\right)^{2}\frac{\partial^{2} f_{0}}{\partial k_{a}\partial k_{b}}\partial_{a}T\partial_{b}T.
\end{aligned}
\label{f1}
\end{equation}

\makeatletter
\renewcommand{\theequation}{B\arabic{equation}}
\renewcommand{\thefigure}{B\arabic{figure}}
\renewcommand{\thetable}{B\arabic{table}}
\section{Magnon relaxation time $\tau_\text{mag}$}\label{MRT}
The magnon relaxation time $\tau_\text{mag}$ can be determined through Fermi golden rule given in Eq. (\ref{tau}). For simplicity, we divide $\tau_\text{mag}$ into
\begin{equation}
\frac{1}{\tau_\text{mag}\left(\mathbf{k}\right)}=\frac{1}{\tau_\text{mag}^{+}\left(\mathbf{k}\right)}+\frac{1}
{\tau_\text{mag}^{-}\left(\mathbf{k}\right)},
\label{APP-B-Tau}
\end{equation}
with
\begin{equation}
\begin{aligned}
\frac{1}{\tau_\text{mag}^{+}\left(\mathbf{k}\right)}=\sum_{\mathbf{k}^{\prime}}W_\text{abs}(\mathbf{k}^{\prime}\mid\mathbf{k})
[1-f\left(\mathbf{k}^{\prime}\right)],\\
\frac{1}{\tau_\text{mag}^{-}\left(\mathbf{k}\right)}=\sum_{\mathbf{k}^{\prime}}W_\text{emit}(\mathbf{k}^{\prime}\mid\mathbf{k})
[1-f\left(\mathbf{k}^{\prime}\right)],
\end{aligned}
\end{equation}
where $W_\text{abs}$ and $W_\text{emit}$ are given in Eq. (\ref{Wae}). Substituting formulas of $\mid\sigma\rangle=\mid\alpha\rangle$ ($\mid\sigma^{\prime}\rangle=\mid\theta\rangle$) determined by Eq. (\ref{alpha}) into Eq. (\ref{Wae}), we have
\begin{eqnarray}
W_\text{abs}&=&\frac{2\pi}{\hbar}j^{2}_\text{ex}n_{\mathbf{k}^{\prime}-\mathbf{k}}\cos^{2}\frac{\alpha+\phi}{2}
\sin^{2}\frac{\theta+\phi}{2}\nonumber\\
&\times&\delta\left(\epsilon_{\mathbf{k}^{\prime}}-\epsilon_\mathbf{k}-\hbar\omega_{\mathbf{k}^{\prime}-\mathbf{k}}\right),\nonumber \\
W_\text{emit}&=&\frac{2\pi}{\hbar}j^{2}_\text{ex}\left(n_{\mathbf{k}-\mathbf{k}^{\prime}}+1\right)\sin^{2}\frac{\alpha+\phi}{2}\cos^{2}
\frac{\theta+\phi}{2}\nonumber\\
&\times&\delta\left(\epsilon_{\mathbf{k}^{\prime}}-\epsilon_\mathbf{k}-\hbar\omega_{\mathbf{k}-\mathbf{k}^{\prime}}\right).
\label{Wae2}
\end{eqnarray}
For a two dimensional case, $\sum_{\mathbf{k}^{\prime}}=\frac{A}{\left(2\pi\right)^{2}}\int d\mathbf{k}^{\prime}$, where $A$ is the area of sample. Here, we introduce a polar coordinate $\left(\alpha,k_{1}\right)$ in which the original point is located at Dirac cone point namely  $\mathbf{k}_{0}=m(\cos\phi,-\sin \phi)/v_{F}\hbar$ [see Fig. \ref{USE}(d)], yielding
\begin{equation}
\begin{aligned}
\mathbf{k}&=\left(k_{1}\cos\alpha+\frac{m\cos\phi}{v_{F}\hbar},k_{1}\sin\alpha-\frac{m\sin\phi}{v_{F}\hbar}\right).\\
\end{aligned}
\end{equation}
Thus, the energy eigenvalues in Eq. (\ref{E-eig}) and the integrated of $d\mathbf{k}$ can be rewritten, respectively, as
\begin{equation}
\begin{aligned}
&\epsilon_{\mathbf{k}}=\epsilon_{k_{1}}=nv_{F}\hbar k_{1},\\
&\epsilon_{\mathbf{k}}-\epsilon_{F}=nv_{F}\hbar\Delta k,\\
&\int d\mathbf{k}\longrightarrow\int d\mathbf{k}_{1}=\int d\alpha \int k_{1}dk_{1}\approx k_{F}\int d\alpha\int d\Delta k,
\end{aligned}
\end{equation}
where $\Delta k$ is measured from $k_{F}$ (Fermi wavenumber). In the following, we take the conduction band as an example, namely $n=1$,  and consider the scattering from the position $\alpha$ to $\theta$ as shown in Fig. ~\ref{2}(d), thus
\begin{equation}
\begin{aligned}
\frac{1}{\tau^{+}_\text{mag}\left(\mathbf{k}\right)}&=\frac{1}{\hbar}\frac{j^{2}_\text{ex}A}{2\pi}k_{F}\int d\theta\int dk^{\prime}_{1}\cos^{2}\frac{\alpha+\phi}{2}\sin^{2}\frac{\theta+\phi}{2}\\
&\times\frac{1}{\exp(\hbar \omega \beta)-1}\left(1-\frac{1}{\exp\left[\beta\left(v_{F}\hbar k_{1}-\epsilon_{F}\right)\right]+1}\right)\\
&\times\delta\left[v_{F}\hbar\left(k^{\prime}_{1}-k_{1}\right)-\hbar\omega\right]\\
&=\frac{1}{\tau^\text{m}_{F}}\int d\theta \frac{Q^{+}_\text{mag}\left(\Delta k_{1}\right)}{\exp\left(\hbar\omega\beta\right)-1} \cos^{2}\frac{\alpha+\phi}{2}\sin^{2}\frac{\theta+\phi}{2}
\end{aligned}
\end{equation}
with
\begin{equation}
\begin{aligned}
Q^{+}_\text{mag}\left(\Delta k_{1}\right)=&\int dk^{\prime}_{1} \left(1-\frac{1}{\exp\left[\beta\left(v_{F}\hbar k^{\prime}_{1}-\epsilon_{F}\right)\right]+1}\right)\\
&\times\delta\left(k_{1}^{\prime}-k_{1}-\frac{\omega}{v_{F}}\right)\\
=&1-\frac{1}{\exp\beta\left[\left(\hbar\omega+\hbar v_{F}\Delta k\right)\right]+1}
\end{aligned}
\label{APP-B-Tau-1}
\end{equation}
and $\frac{1}{\tau^\text{m}_{F}}=\frac{k_{F}}{2\pi}\frac{j^{2}_\text{ex}A}{v_{F}\hbar^{2}}$. $\hbar\omega$ corresponds to the magnon energy with $2k_{F}\sin\left(\theta-\alpha\right)$ wavenumber.
Similarly, one can have
\begin{equation}
\begin{aligned}
\frac{1}{\tau^{-}_\text{mag}\left(\mathbf{k}\right)}&=\frac{1}{\tau^\text{m}_{F}}\int^{2\pi}_{0} d\theta \sin^{2}\frac{\phi+\alpha}{2}\cos^{2}\frac{\phi+\theta}{2}\\
&\times\!\left(\frac{1}{\exp\left(\hbar\omega\beta\right)-1}+1\right)\!\frac{\exp\beta\left(\hbar v_{F}\Delta k-\hbar\omega\right)}{\exp\beta\left(\hbar v_{F}\Delta k-\hbar\omega\right)+1}.
\end{aligned}
\label{APP-B-Tau-2}
\end{equation}
Based on Eqs. (\ref{APP-B-Tau}), (\ref{APP-B-Tau-1}) and (\ref{APP-B-Tau-2}), we obtain
\begin{equation}
\begin{aligned}
\frac{1}{\tau_\text{mag}\left(\mathbf{k}\right)}&=\frac{1}{\tau_\text{mag}\left(\alpha,\Delta k\right)}\\
&=\frac{1}{\tau^\text{m}_{F}}\int^{2\pi}_{0} d\theta V_\text{mag}\left(\theta+\phi,\alpha+\phi,\Delta k\right)
\end{aligned}
\label{APP-B-Tau-4}
\end{equation}
with the integrand $V_\text{mag}\left(\theta+\phi,\alpha+\phi,\Delta k\right)$ defined as
\begin{equation}
\begin{aligned}
&V_\text{mag}\left(\theta+\phi,\alpha+\phi,\Delta k\right)\\
&=\cos^{2}\left(\frac{\alpha+\phi}{2}\right)\sin^{2}\left(\frac{\theta+\phi}{2}\right)
V^{+}_\text{mag}\left(\theta,\alpha,\Delta k\right)\\
&+\sin^{2}\left(\frac{\alpha+\phi}{2}\right)\cos^{2}\left(\frac{\theta+\phi}{2}\right)V^{-}_\text{mag}\left(\theta,\alpha,\Delta k\right),\\
\end{aligned}
\label{APP-B-V}
\end{equation}
where
\begin{eqnarray}
V^{+}_\text{mag}\left(\theta,\alpha,\Delta k\right)&=&\frac{1}{e^{\beta\hbar\omega}-1}\left(1-\frac{1}{e^{\beta\left(\hbar v_{F}\Delta k+\hbar\omega\right)}+1}\right),\nonumber\\
V^{-}_\text{mag}\left(\theta,\alpha,\Delta k\right)&=&\left(1-\frac{1}{e^{\beta\left(\hbar v_{F}\Delta k-\hbar\omega\right)}+1}\right)\nonumber\\
&\times&\left(\frac{1}{e^{\beta\hbar\omega}-1}+1\right).
\label{APP-B-VPM}
\end{eqnarray}

\makeatletter
\renewcommand{\theequation}{C\arabic{equation}}
\renewcommand{\thefigure}{C\arabic{figure}}
\renewcommand{\thetable}{C\arabic{table}}
\section{The formula of $\Delta S$ induced by the magnon asymmetry scattering in heterostructures TI/MTI}\label{expresion-USE}
Taking the magnetic relaxation time in Eq. (\ref{tau1}) into Eq. (\ref{alph-mag}), we can have $\alpha^{(2)}_{xx,\text{mag}}$ in polar coordinates ($\alpha,\Delta k$) (see Appendix \ref{MRT} for detail), where $\alpha$ is polar angle measured from $k_{x}$ axis and $\Delta k$ is the radius measured from $k_{F}$ (Fermi momentum), as
\begin{widetext}
\begin{equation}
\begin{aligned}
\alpha^{(2)}_{xx,\text{mag}}&=\frac{2e\left(\tau^{0}\right)^{3}}{\tau^\text{m}_{F}}k_{F}\int d\Delta k\int d\alpha \int d\theta V_\text{mag}\left(\theta+\phi,\alpha+\phi,\Delta k\right)\left[v^{2}_{x}\frac{v_{F}\Delta k}{T^{2}}\frac{\partial f_{0}}{\partial k_{x}}+v_{x}\left(\frac{v_{F}\Delta k}{ T}\right)^{2}
\frac{\partial^{2} f_{0}}{\partial^{2} k_{x}}\right].
\end{aligned}
\label{App-C-alph-mag}
\end{equation}
\end{widetext}
For $v_{x}\left(\mathbf{k}\right)$,
\begin{equation}
\begin{aligned}
v_{x}&=\frac{1}{\hbar}\frac{\partial\epsilon_{\mathbf{k}}}{\partial k_{x}}=v_{F}\cos\alpha.
\end{aligned}
\end{equation}
For $\frac{\partial f_{0}}{\partial k_{x}}$,
\begin{equation}
\frac{\partial f_{0}}{\partial k_{x}}\left(\alpha,\Delta k\right)=-\frac{P}{\left(P+1\right)^2}\beta v_{F}\hbar \cos\alpha.
\end{equation}
For $\frac{\partial^{2} f_{0}}{\partial k_{x}^{2}}$,
\begin{equation}
\begin{aligned}
\frac{\partial^{2} f_{0}}{\partial k_{x}^{2} }\left(\alpha,\Delta k\right)=&-\frac{\beta\hbar v_{F}}{k_{F}}\frac{P}{\left(P+1\right)^2}\sin^{2}\alpha+\left(\beta\hbar v_{F}\right)^{2}\cos^{2}\alpha\\
&\times\frac{P\left(P-1\right)}{\left(P+1\right)^{3}}\\
\approx& \left(\beta\hbar v_{F}\right)^{2}\frac{P\left(P-1\right)}{\left(P+1\right)^{3}}\cos^{2}\alpha,
\end{aligned}
\label{APP-C-P-f2}
\end{equation}
with $P=e^{\beta \hbar v_{F}\Delta k}$. Here, in the third line of Eq. (\ref{APP-C-P-f2}), we can ignore the first term since  $\beta\hbar v_{F}k_{F}\gg 1$ except for the immediate vicinity of the Dirac point. Meanwhile,
\begin{equation}
\begin{aligned}
\sigma_{xx}&=-\frac{e^{2}}{\hbar}\frac{1}{\left(2\pi\right)^{2}}\int d\mathbf{k}\tau \left(\mathbf{k}\right)v_{x}\left(\mathbf{k}\right)\frac{\partial f}{\partial k_{x}}\\
&\approx-\frac{e^{2}}{\hbar}\frac{\tau^{0}k_{F}}{\left(2\pi\right)^{2}}\int d\Delta k\int^{2\pi}_{0} d\alpha v_{x}\left(\alpha\right)\frac{\partial f}{\partial k_{x}}\left(\Delta k,\alpha\right)\\
&=\frac{e^{2}}{4\pi\hbar}v_{F}k_{F}\tau^{0}.
\end{aligned}
\label{APP-B-Con}
\end{equation}
Therefore, from  Eq. (\ref{DeltaS}) we can determine the expression of $\Delta S$ for characterizing the USE as
\begin{equation}
\begin{aligned}
{\Delta S}=-&\zeta\frac{A \epsilon_{F}\Delta T}{lT}
\!\int\! d\Delta k\int\! d\alpha\!\int\! d\theta V_\text{mag}\left(\theta+\phi,\alpha+\phi,\Delta k\right)\\
&\times\left(\frac{\hbar v_{F}}{k_\text{B}T}\right)^{2}\Delta k\frac{P\cos^{3}\alpha }{\left(P+1\right)^2} \left(1+
\frac{1-P}{P+1}\frac{\hbar v_{F}\Delta k}{k_\text{B}T}\right),
\end{aligned}
\label{App-C-DeS-F}
\end{equation}
where $\zeta=4k_\text{B}j^{2}_\text{ex}\left(\tau^{0}\right)^{2}/(e\hbar^{3}v_{F})$. Let $x=\hbar v_{F}\Delta k/ (k_\text{B}T)$ in Eq. (\ref{App-C-DeS-F}) and taking the formula of $P=e^{\hbar v_{F}\Delta k/ (k_\text{B}T)}$ into it, we have obtained the expression of $\Delta S$ in Eq. (\ref{DeS-F}) in the main text.  In the following, we consider a situation in which we neglect magnon dispersion and use $g\mu_{B}B$ as magnon energy, namely $\hbar\omega=g\mu_{B}B$. Thus, $V^{+(-)}_\text{mag}\left(\theta,\alpha,\Delta k\right)$ in Eq. (\ref{APP-B-VPM}) will be independent of $\theta$ and $\alpha$ and thus, we rewrite them as $V^{+(-)}_\text{mag}\left(\Delta k\right)$. Therefore, the integration of the angle-dependent part in Eq. (\ref{App-C-DeS-F}) is found to be
\begin{equation}
\begin{aligned}
\Xi=&\int^{2\pi}_{0}d\alpha\int^{2\pi}_{0}d\theta V_\text{mag}\left(\theta+\phi,\alpha+\phi,\Delta k\right)\cos^{3}\alpha\\
=&\int^{2\pi}_{0}d\alpha\int^{2\pi}_{0}d\theta \cos^{3}\alpha\left[\cos^2\frac{\phi+\alpha}{2}\sin^2(\frac{\phi+\theta}{2})\right.\\
&\left.\times V^{+}_\text{mag}\left(\Delta k\right)+\sin^2\frac{\phi+\alpha}{2}\cos^2(\frac{\phi+\theta}{2})V^{-}_\text{mag}\left(\Delta k\right)\right]\\
=&\frac{3\pi^{2}}{8}\left[V^{+}_\text{mag}\left(\Delta k\right)-V^{-}_\text{mag}\left(\Delta k\right)\right]\cos\phi.
\end{aligned}
\label{APP-C-Xi}
\end{equation}
Thus,
\begin{eqnarray}
\frac{\Delta S}{\Delta T}&=&-\zeta\frac{3\pi^{2} Al \epsilon_{F}\Delta T}{8 T}\cos\phi
\int d\Delta k  \left[V^{+}_\text{mag}\left(\Delta k\right)-V^{-}_\text{mag}\left(\Delta k\right)\right]\nonumber\\
&\times&\left(\frac{\hbar v_{F}}{k_\text{B}T}\right)^{2}\Delta k\frac{P\cos^{3}\alpha }{\left(P+1\right)^2} \left(1+
\frac{1-P}{P+1}\frac{\hbar v_{F}\Delta k}{k_\text{B}T}\right)\nonumber\\
&=&\zeta\frac{3\pi^{2} Al \tau^{m}_{F} \epsilon_{F}\Delta T}{8 T}\cos\phi\int dx \left(\frac{1}{\tau^{+}}-\frac{1}{\tau^{-}}\right)\nonumber\\
&\times&\frac{xe^{x}}{\left(1+e^{x}\right)^{2}}\left(1+\frac{1-e^{x}}{e^{x}+1}x\right).
\label{App-C-DeS-F-1}
\end{eqnarray}
To obtain the third line, we have make use of $x=\hbar v_{F}\Delta k/ (k_\text{B}T)$ and $1/\tau^{+(-)}=V^{+(-)}_\text{mag}/\tau^\text{m}_{F}$ given in Eqs. (\ref{taup}) and (\ref{taum}).

\makeatletter
\renewcommand{\theequation}{D\arabic{equation}}
\renewcommand{\thefigure}{D\arabic{figure}}
\renewcommand{\thetable}{D\arabic{table}}

\section{The analysis of symmetries (or parties) of $1/\tau_\text{mag}$ and the $\cos\phi$-dependence of magnetization of USE}\label{symmetry-cos}

In this appendix, the symmetries (or parties) of $1/\tau_\text{mag}$ and the $\cos\phi$-dependence of ${\Delta S}/{\Delta T}$ in Eq. (\ref{non-Se}) will be analyzed. Before detail analysis, we would like to introduce two concepts: 1) it's noted that the symmetry /antisymmetry of a function $f\left(k_{x},k_{y}\right)$ corresponds to even/ odd parties with respect to $k_{x}$, respectively, namely $f\left(k_{x},k_{y}\right)=f\left(-k_{x},k_{y}\right)$ /$f\left(k_{x},k_{y}\right)=-f\left(-k_{x},k_{y}\right)$. This is a consequence of mirror symmetry with respect $k_{y}-k_{z}$ plane in momentum space. 2) In polar coordinate ($\alpha,\Delta k$), where $\alpha$ is polar angle measured from $k_{x}$ axis and $\Delta k$ is the radius measured from Fermi momentum $k_{F}$, $\cos\alpha$ ($\sin\alpha$) dependence of function $f$  represents that $f$ is odd (even) function with respect to $k_{x}$, respectively.

To understand how $\cos\phi$-dependence of ${\Delta S}/{\Delta T}$ in Eq. (\ref{non-Se}) appears physically, we can investigate symmetry of magnon scattering time with respect to $k_{x}$. As mentioned above Eq. (\ref{APP-C-Xi}), when neglecting magnon dispersion and use $gu_\text{B}B$ as magnon energy, $V^{+(-)}_\text{mag}\left(\theta,\alpha,\Delta k\right)$ in Eq. (\ref{APP-B-VPM}) will be independent of $\theta$ and $\alpha$ and is rewritten as $V^{+(-)}_\text{mag}\left(\Delta k\right)$. Hence, the magnon scattering time $\tau_\text{mag}$ in Eq. (\ref{APP-B-Tau-4}) is found to be
\begin{equation}
\begin{aligned}
\frac{1}{\tau_\text{mag}(\alpha,\Delta k)}=&\int^{2\pi}_{0}\frac{d\theta}{\tau^\text{m}_{F}} V_\text{mag}\left(\theta+\phi,\alpha+\phi,\Delta k\right)\\
=&\int^{2\pi}_{0}\frac{d\theta}{\tau^\text{m}_{F}}\left[\cos^2\frac{\phi+\alpha}{2}\sin^2(\frac{\phi+\theta}{2})V^{+}_\text{mag}\left(\Delta k\right)\right.\\
&\left. +\sin^2\frac{\phi+\alpha}{2}\cos^2(\frac{\phi+\theta}{2})V^{-}_\text{mag}\left(\Delta k\right)\right]\\
=&\frac{1}{\tau^\text{S}_\text{mag}}+\frac{1}{\tau^\text{A}_\text{mag}}
\end{aligned}
\label{APP-D-Xi}
\end{equation}
with
\begin{equation}
\begin{aligned}
\frac{1}{\tau^\text{S}_\text{mag}}=&\frac{\pi}{2\tau^\text{m}_{F}}\left[\left(V^{+}_\text{mag}\left(\Delta k\right)+V^{-}_\text{mag}\left(\Delta k\right)\right)\right.\\
&\left.-\left(V^{+}_\text{mag}\left(\Delta k\right)-V^{-}_\text{mag}\left(\Delta k\right)\right)\sin\phi\sin\alpha\right],\\
\frac{1}{\tau^\text{A}_\text{mag}}=&\frac{\pi}{2\tau^\text{m}_{F}}
\left(V^{+}_\text{mag}\left(\Delta k\right)-V^{-}_\text{mag}\left(\Delta k\right)\right)\cos\phi\cos\alpha,
\end{aligned}
\label{APP-D-AS}
\end{equation}
where the superscript ``{A}" (``{S}") in $\tau^\text{A}_\text{mag}$ ($\tau^\text{S}_\text{mag}$) refer to antisymmetry (symmetry), respectively. ${1}/{\tau^\text{S}_\text{mag}}$ ( ${1}/{\tau^\text{A}_\text{mag}}$) gives the symmetry (antisymmetry) part of ${1}/{\tau_\text{mag}}$ when reversing $k_{x}$, respectively, namely ${1}/{\tau^\text{S}_\text{mag}(k_{x},k_{y})}={1}/{\tau^\text{S}_\text{mag}(-k_{x},k_{y})}$ and ${1}/{\tau^\text{A}_\text{mag}(k_{x},k_{y})}=-{1}/{\tau^\text{A}_\text{mag}(-k_{x},k_{y})}$. Exploiting the parity of Table.\ref{parities}, one can find that the term in brackets in Eq. (\ref{alph-mag}) is a odd function of $k_{x}(\propto\cos\phi)$. Therefore, only the antisymmetric part of ${1}/{\tau_\text{mag}}$ gives finite value to $a^{\left(2\right)}_{xx,\text{mag}}$, namely USE induced by the asymmetry magnon scattering arising from ${1}/{\tau^\text{A}_\text{mag}}$.

The $\cos\phi$-dependence of ${1}/{\tau^\text{A}_\text{mag}}$ [Eq. (\ref{APP-D-AS})] gives rise to cosine dependence on magnetization angle, namely only $m_{y}(\propto \cos\phi$) part of magnetization has contribution to the unidirectional Seebeck effect. The $m_{x}(\propto \sin\phi$) part of magnetization induce the symmetric magnon scattering [${1}/{\tau^\text{S}_\text{mag}}$]  which can not lead to the antisymmetric contribution in Seebeck effect when reversing the temperature gradient.

\begin{figure}[tb]
\centering
\includegraphics[width=1.0\linewidth]{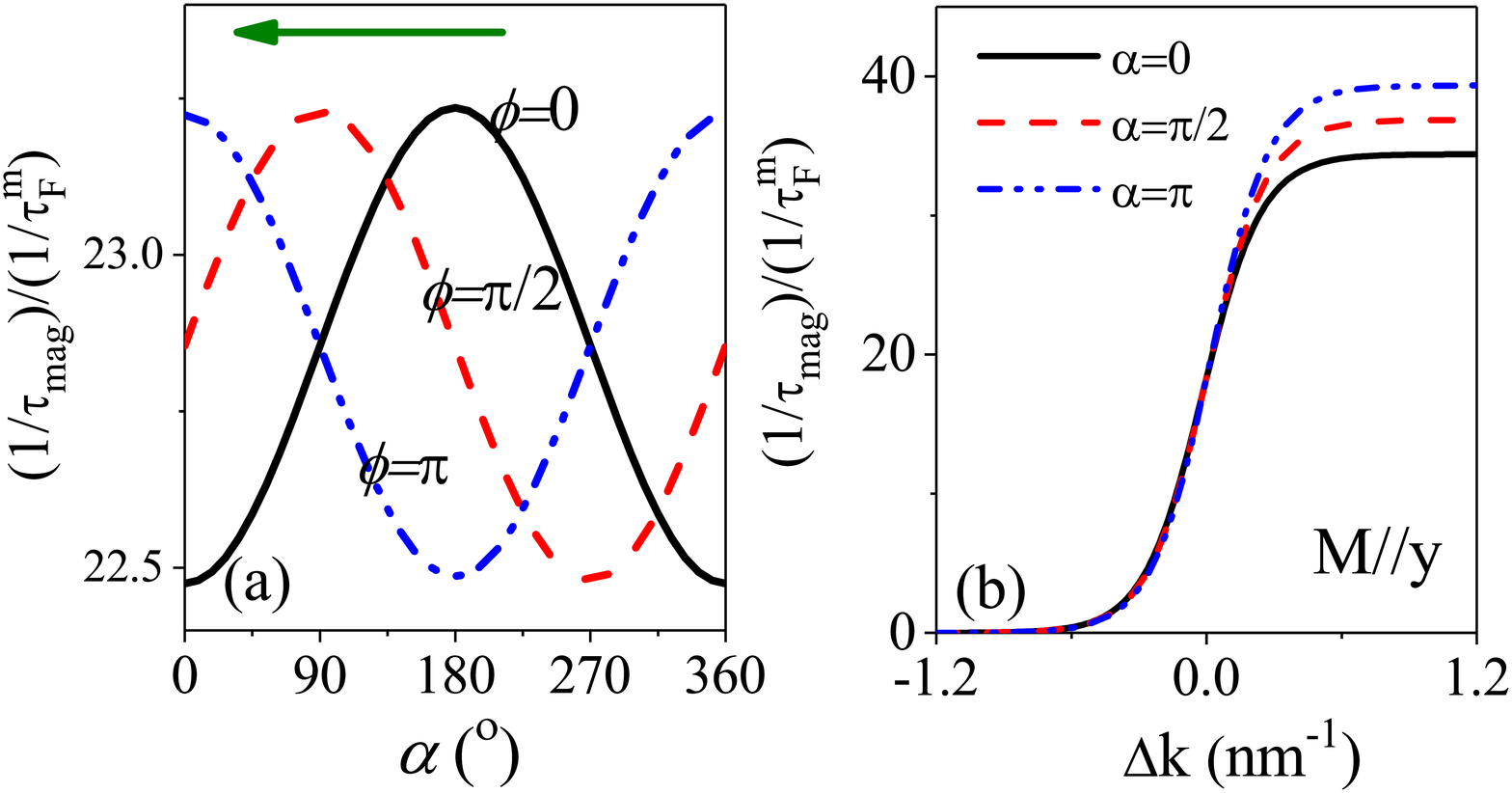}
\caption{$\alpha$ dependence of $1/\tau_\text{mag}$ for different $\phi$ is shown in (a) and $ \Delta k$ dependence of $1/\tau_\text{mag}$ for different polar angle $\alpha$ is shown in (b) with magnetization along $y$ direction, namely $\phi=0$. The magnetic field is $B=1$ T and the temperature is T=10K. $\alpha$ is polar angle measured from $k_{x}$ axis and $\Delta k$ is the radius measured from Fermi momentum $k_{F}$. The azimuth angle $\phi$ is used to indicate the magnetization orientation measured from the $y$ axis.  }
\label{APP-c-tau}
\end{figure}

Figure \ref{APP-c-tau}(a) shows the variation of $1/\tau_\text{mag}$ with $\alpha$  for different magnetization orientation (i. e., $\phi$). One can identify $1/\tau_\text{mag}$ would be expressed as the function of ($\sin\alpha,\cos\alpha)$ [given in Eqs.(\ref{APP-D-Xi})(\ref{APP-D-AS})]. When increasing $\phi$, the whole curve will shift towards lower $\alpha$. The parities of $1/\tau_\text{mag}$ is strongly influenced by the magnetization orientation. Since the absence of even properties in magnon relaxation time guarantees the existence of unidirectional Seebeck effect. From Fig. \ref{APP-c-tau}(a), one can observe that only when the magnetization is aligned to $x$-direction (i. e., $\phi=\pi/2$ or $\phi=3\pi/2$ (not shown)),  $1/\tau_\text{mag}$ is symmetric with respect to the mirror plane of $k_{y}-k_{z}$ ($\alpha=90^{0},180^{0}$), namely the presence of even properties about $k_{x}$. Thus, when magnetic field is not aligned $x$-direction, the proposed effect would be observed.

\end{document}